# Use of NMR to Test Molecular Mobility during Chemical Reaction

*Huan Wang[a,,b,1] Tian Huang,[a,1] and Steve Granick[a,c,*]*

[a] *Center for Soft and Living Matter, Institute for Basic Science (IBS), Ulsan 44919, South Korea*

[b] *College of Chemistry and Molecular Engineering, Peking University, Beijing, 100871, P. R. China*

[c] *Departments of Chemistry and Physics, Ulsan National Institute of Science and Technology (UNIST), Ulsan 44919, South Korea*

* Corresponding author email: sgranick@gmail.com

[1] these authors contributed equally

Abstract

We evaluate critically the use of pulsed gradient spin-echo nuclear magnetic resonance (PGSE NMR) to measure molecular mobility during chemical reactions. With raw NMR spectra available in a public depository, we confirm boosted mobility during the click chemical reaction (*Science* **2020**, 369, 537) regardless of the order of magnetic field gradient (linearly-increasing, linearly-decreasing, random sequence). We also confirm boosted mobility for the Diels-Alder chemical reaction. The conceptual advantage of the former chemical system is that constant reaction rate implies constant catalyst concentration, whereas that of the latter is the absence of a paramagnetic catalyst, precluding paramagnetism as objection to the measurements. Data and discussion in this paper show the reliability of experiments when one avoids convection, allows decay of nuclear spin magnetization between successive pulses and recovery of its intensity between gradients, and satisfies quasi-steady state during the time window to acquire each datum. Especially important is to make comparisons on the time scale of actual chemical reaction kinetics. We discuss possible sources of mistaken conclusions that are desirable to avoid.

MAIN TEXT

A surge of interest to understand moving entities that consume energy during the course of their motion, so-called 'active matter',[1-7] is entering a phase that goes beyond earlier focus on colloidal and nanoparticle active mobility, and today considers the role of molecules as active matter.[3-4, 7-9] We consider here pulsed gradient spin-echo (PGSE) NMR.[10-14] The technique often is combined with diffusion ordered NMR spectroscopy (DOSY) analysis in which one dimension represents chemical shift data while the second dimension resolves species by their diffusion properties.[15-19] This technique presents many attractive features. Unlike fluorescence-based methods, it does not require chemical tags on the sample. Unlike dynamic light scattering, there is no minimum molecular size[20-21] It simultaneously identifies different chemical species and their abundance in the same sample. It can be extended to out-of-equilibrium situations.[18-19, 22-24] For example, it can discriminate reactive intermediates of organometallics, determining their aggregation number, solvation states and the identification of new reaction intermediates based on the unique capability of this method to correlate structure and mobility.[25-26]

Extending a recent study from this laboratory,[27] and our response[28] to a critical comment,[29] we evaluate here critically the soundness of the PGSE method to assess molecular mobility in common chemical reactions. As precautions needed to avoid convection are discussed amply in the literature[30-31] we do not discuss them. Discussion in this paper dwells on two issues. We discuss the conditions to satisfy, while acquiring each datum, the quasi-steady-state condition in this out-of-equilibrium condition. Secondly, we discuss changes in the relaxation time of nuclear spin magnetization parallel to the external magnetic field that could lead to a change of signal intensity, a matter that is testable by varying the diffusion delay time.[12-13] Finally, we highlight that to check physical reasonability of conclusions, mobility measurements should be compared to the time scale of actual chemical reaction kinetics.

The concept is summarized in Fig. 1. The liquid sample is mounted in an NMR tube. After a pulse magnetic field to align nuclear spins of the chemical moieties of interest, a linear magnetic field gradient along the cylinder length encodes spatial information. After a waiting time during which the chemical moieties of interest experience self-diffusion, the signal is negated by pulses that recover the original nuclear spins – the only signal left is from chemical moieties that migrated to a different vertical location in the field gradient. These procedures can be accomplished various ways; we have adopted one of the standard methods, to apply two subsequent echo pulses, the exact reverse of the first pulse. In the concept of this measurement, the above procedure is repeated, each time with a different magnetic field gradient. Attenuated intensity in the recovered signal, plotted against the gradient field squared in these multiple experiments, gives using standard analysis a number proportional to the self-diffusion coefficient. Given the claim that gradients composed of linearly-increasing, linearly-decreasing and random sequences give inconsistent results,[29] here we compare such findings with measurements made in our laboratory. Our favorable comparison between the three procedures allows one to quantitatively assess the quasi-steady-state assumption on which reliable measurement depends.

_Aqueous click reaction_. Fig. 2 shows data averaged from multiple independent experiments. Representative raw data from individual experiments is freely available from a public depository[32]. With chemical reaction conditions selected to be the same as in our original report,[27] the aqueous click reaction between alkyne and azide (Fig. 2A) was allowed to proceed at room temperature with reactant and catalyst concentrations that produce constant reaction rate for the first 80 min and reaction completion at 120 min (Fig. 2B and 2C), significantly longer than the 3-5 min needed to acquire each datum. Chemical shifts of relevant NMR peaks are cleanly resolved (Fig. 2D). For the conditions specified[27] the relative change of peak intensity in one scan is 0.2%, while to produce the magnetic field gradient (16 scans) it is at most 3%, small relative to the 5% - 95% imposed magnetic field gradient intensity. The assertion that concentrations change substantially while measuring each datum[29] is not supported.

Fig. 2E shows data comparing findings obtained using three different sequences of magnetic field -- linearly decreasing, random and linearly increasing -- regarding the alkyne reactant at chemical shift 2.9 (peak 1, e-h panels) and 4.2 (peak 2, a-d panels) ppm (parts-per-million). The curves obtained from random sequences (c, g panels) are more scattered than for linearly increasing (d, h panels)[27] and linearly decreasing sequences (b, f panels) and we speculate that hardware may influence this as the NMR instrument was designed and calibrated for linear sequences, whereas we randomized the sequence using a non-uniform sampling method. Notice that two chemical moieties on the same reactant display quantitative differences, though both show differences; intermediates of the chemical reaction may also contribute in ways that presently are imperfectly understood, as we noted previously.[27] Overall, the ratio of apparent diffusion coefficient obtained by random sequence measurement, to that from linearly increasing sequence measurement, is unity within experimental uncertainty at all reaction times (a,e panels). One also notices a tendency for persistently slightly smaller diffusion coefficient for alkyne peak at 4.2 and azide peak at 3.8 ppm during the first 50 min, perhaps reflecting concentration consumption during this interval of most rapid reaction rate (Fig. 2C). This was not observed for the ligand peak 10, which is not consumed by this chemical reaction.

It is claimed that changes of catalyst concentration during a chemical reaction can modify the relaxation time of reactants so as to contribute significantly as non-diffusive signal attenuation in diffusion measurement,[29] but this cannot be the case for either catalyst molar abundance nor for abundance of catalyst in the active oxidation state, as reaction rate was linear during the first 80 min of data shown here. This signifies that the amount of active catalyst Cu(I) was constant, so the remaining Cu(II) if any in the reaction should be constant as well.

*Comparison to contrary assertions in reference 29*. For Fig. 3, the raw NMR spectra are freely available from a public depository[32]. But raw data for reference 29 are not available publicly, to the best of our knowledge.

From the perspective of considering hypothetically how mistaken interpretations might emerge, consider the raw data we posted online. Regarding parameters optimized to measure the alkyne and aide peaks, the delay time was long, 10 s, and the pulse length was long, 1100 μs. Optimal pulse parameters differ for the ascorbate and water peaks for the reasons discussed in the next paragraph. It would be mistaken to analyze those peaks with these unsuitable parameters; for example, to use this pulse length gives, at the extremes of 5% and 95% magnetic field gradient, intensity differences prohibitively too different to give meaningful results. If instrumental parameters do not give linear attenuation in the relevant range of measurements, conclusions from analyzing the resulting experiments will not be meaningful. The same experimental parameters should not be used to analyze all peaks.

From this perspective, we call attention to relevant experimental parameters. Among the most important are to optimize the parameters that determine signal attenuation: gradient strength, pulse duration, and diffusion time. Normally we kept constant the duration and diffusion time (cf. Table S2 in ref. 27) and varied gradient strength to obtain diffusion coefficient from fits to the Stejskal-Tanner equation (Fig. 1C). For this purpose, the magnetic gradient should be applied over a distance such that at the extremes of gradient (e.g. 5% and 95%) one observes a peak decay proportionally, for example 19 times difference. We find the optimal pulse duration (p30, Bruker NMR) to be approximately 800 μs for water and 1100-1200 μs for alkyne and azide, when diffusion time ($D_2O$) is set as 50 ms.

Hypothetically, we also note another possible origin of mistaken conclusions, which is to ignore reaction kinetics. If one compares data under conditions where the chemical reaction continues for the extended time of 150 min,[29] it is not meaningful to compare diffusion over just the first 50 min of reaction, but this is the comparison described in reference 29. In our laboratory, the chemical reaction extended to 150 min only for lower catalyst concentrations than those asserted in ref. 29. This discrepancy encourages us to specify here our reaction protocol when mixing reagents and catalyst: we added ascorbate as the final step in solid form just before reaction was initiated. The reason is that this reducing agent reduces Cu(II) to Cu(I), which is the active catalyst form. If data in ref. 29 was obtained from using stock solutions, then the actual catalyst concentration would be less than the nominal value as reducing agent is air sensitive and can be oxidized by oxygen.

By lowering the catalyst concentration below the nominal value reported in reference 29, we reproduced the slower kinetics reported in reference 29. With detailed reaction conditions specified in the caption of Fig. 3, we then performed diffusion NMR measurements using random PFSE NMR pulse sequences recommended in ref. 29. As ref. 29 asserted that boosted diffusion might originate from choosing an unsuitably-rapid relaxation delay time, we draw to the fact that these experiments employed the long relaxation delay time of 10 s, which is 5 times longer than the slowest spin-relaxation time (2 s) asserted in ref. 29.

One observes in Fig. 3A (alkyne reactant) and Fig. 3B (azide reactant) that reactant diffusion was faster during reaction than after its completion, by factors of 1.2 to 1.3. Fig. 3C shows explicitly the kinetics of this reaction, reactant and product reactions plotted against time. Fig. 3D shows the reaction rate plotted against time – this number was nearly constant for 150-200 min, consistent with statements in ref. 29. It is evident that boosted diffusion was observed during the chemical reaction regardless of all these considerations.

*Diels-Alder reaction in acetonitrile*. The reaction rate, inferred from conventional one-dimensional NMR, is plotted in Fig. 4B. The proton peak assignments in $^1$H-NMR are shown in Fig. 4C. Fig. 4D shows the excellent agreement obtained between magnetic field gradients produced three different ways – linearly-increasing, linearly-decreasing, random. The ratio of apparent diffusion coefficient obtained by random sequence measurement, to that from linearly increasing sequence measurement, is unity within experimental uncertainty at all reaction times (panel a). Plotting explicitly the apparent diffusion normalized by that at the end of reaction against reaction time using the three methods, the raw data agree for linearly-decreasing magnetic field gradient (panel b), random magnetic field gradient (panel c), and linearly-increasing random field gradient (panel d). Within the experimental uncertainty (5%, 4%, 3%), the respective maximum values agree.

The Diels-Alder reaction involves no metal catalyst so an argument based on supposed changes of metal catalyst concentration[29] could not hold for this second example of a chemical system in which PGSE NMR shows boosted diffusion during chemical reaction. Moreover, molecules undergoing chemical reaction may display time-dependent changes in relaxation time, unlike the case for equilibrium systems. A useful safety check is to use progressively longer recovery times (D1) to probe for any systematic artefact and choose the shortest one that leads to credible measurement. In an earlier report we varied this quantity between 1.5 s and 5 s, observing that findings were indistinguishable with 5 s and 10 s delay.[27]

*Other experimental concerns.* Synthesizing many suggestions gleaned from surveying a long literature,[10-14, 18-19, 23-25, 30-31, 33-44] we note that specialists are well aware that other concerns about the PGSE NMR method include the possibility of electrical eddy currents caused by fast switching of the applied gradient pulse,[38-39] gradient field non-uniformity,[40-41] and convection currents caused by temperature gradients[42-43] and chemical reactions one of whose products is a gas.[30, 44] Eddy currents usually are greatly reduced by use of a shielded gradient system but cannot be fully excluded but as eddy currents cannot be eliminated, it is reasonable that they may differ according to the sequence of magnetic field gradients (increasing, decreasing, and random). This may explain why we find greater measurement scatter in the latter case.[15] Systematic error will result to the extent that the magnetic gradient is not well-defined in the sample cylinder direction.

Helpfully for the analysis of data, temperature gradients and gradient field non-uniformity commonly produce deviations from the expected linearity of intensity dependence on gradient field squared (Fig. 1C). Moreover, as convection amplitude increases with sample size, control experiments using sample tubes of different sizes help to check the hypothetical possibility of macroscopic convection in addition to use convection suppression pulses[27, 30-31, 34, 44] as we did previously to carefully choose reaction system for study.[27] Helpfully for identification of the hypothetical possibility that reaction consumption might interfere with satisfying the steady-state assumption needed for data analysis,[15] note that rate of reaction consumption is often linear or power-law according to the reaction rate, unlike the exponential decay due to gradient field, so goodness of fit to the linearity in Fig. 1C serves an additional safety check. In fact, methods like trilinear analysis[17] work to separate this reaction-induced intensity change from diffusion-induced signal attenuation. Measurements are most robust for isolated, highly abundant peaks of well-defined shape, for which we find the standard deviation of reproducible measurements to be 1-2%. For overlapping peaks, we find that the standard deviation can be as high as 30%.[10, 18] Hence, especially when dealing with small effects, one should carefully choose to study peaks of the former type to encourage credible results and reliable comparisons.[27]

*Summary*. Experimental tests summarized in this paper underscore the usefulness of PGSE NMR to study active molecular matter. Summarizing experimental considerations and useful cross-checks, and applying these methods to studying molecular mobility in the click reaction system and in a Diels-Alder chemical reaction system, we confirm the useful experimental information that the order of magnetic field imposition (linearly-increasing, linearly-decreasing, or randomized) is immaterial. The proposition that molecules when they undergo chemical reaction can behave as active matter generalizes the perspective that mechanisms of even the simplest chemical reactions may require rethinking.[45]

**Corresponding Author**

Steve Granick − Center for Soft and Living Matter, Institute for Basic Science, Ulsan 44919, South Korea; Department of Chemistry, Ulsan National Institute of Science and Technology, Ulsan 44919, South Korea;

orcid.org/0000-0003-4775-2202;

email: sgranick@gmail.com

**Author Contributions**

The manuscript was written through contributions of all authors. All authors have given approval to the final version of the manuscript.

**Funding Sources**

This work was supported by the taxpayers of South Korea through the Institute for Basic Science, project code IBS-R020-D1.

**Notes**

The authors declare no competing financial interest.

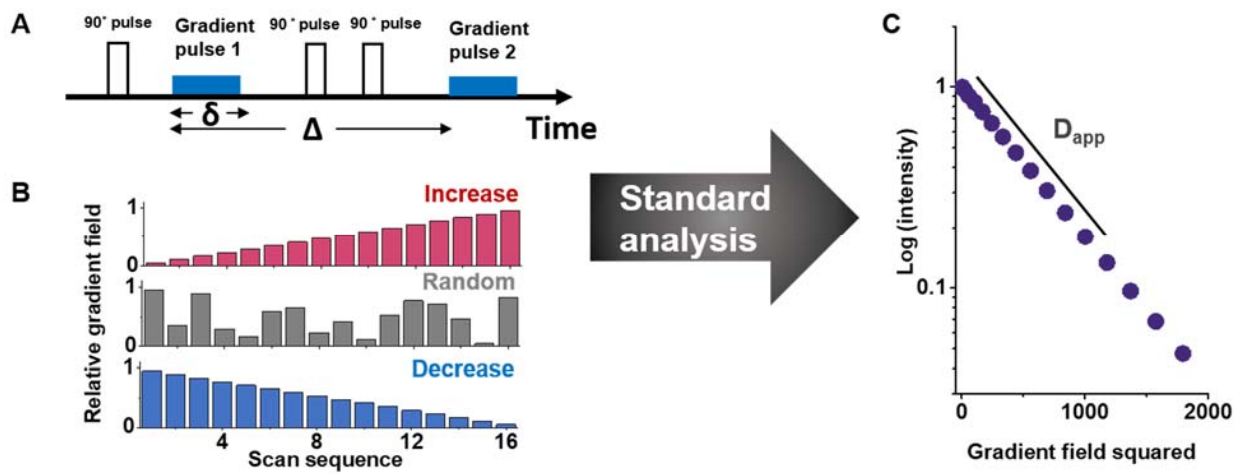

**Figure 1. Schematic concept of PGSE NMR.** The liquid sample is mounted in a conventional cylindrical NMR tube. (A) A standard sequence of pulses and magnetic field gradient is applied to spatially encode spins of interest by dephasing, resulting a reduced signal after an echo pulse is applied. (B) It is standard to set up magnetic gradients with the magnitude of each of them, commonly 16 discrete fields, in linearly-increasing sequence. This paper compares to findings using a random sequence and linearly-decreasing sequence. (C) The self-diffusion coefficient is proportional, by standard analysis, to a plot of logarithmic signal intensity against square magnetic field gradient.

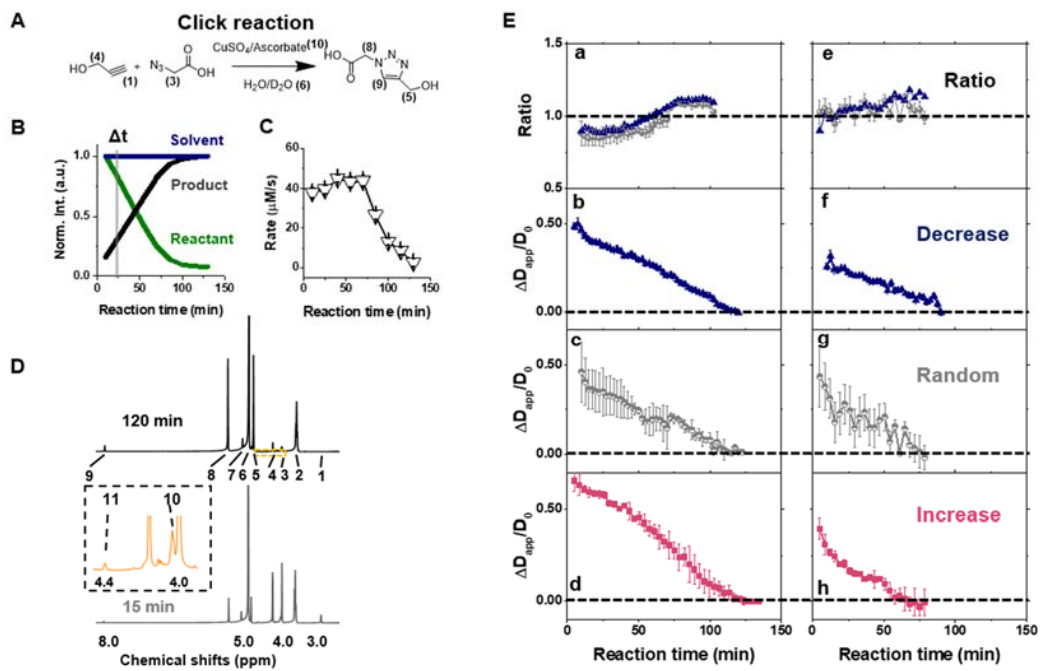

**Figure 2. Diffusion data for the click reaction compared using linearly-increasing, linearly decreasing, and random sequences of magnetic fields. (A)** Chemistry of this reaction. (**B, C**) Kinetics of this reaction, showing smoothed data for the conditions of ref.[27] Grey shaded vertical bar indicates the time scale, 3-5 min, needed for one measurement. (**D**) NMR spectra near the start (15 min) and end (120 min) of reaction, showing the NMR peak assignments. Inset enlarges the catalyst ligand peaks. (**E**) Comparisons of data taken using three sequences of magnetic gradients for reactant peaks, panels a to d for 4.2 ppm, panels e to h for 2.9 ppm. Panels a and e show ratio of diffusion coefficients of chemical moieties obtained from random sequence (grey spheres) and linearly decreasing sequence (navy triangles) to linearly increasing sequence (rectangles). Panels b to h show increased diffusion coefficient, relative to $D_o$ at the end of reaction, plotted against reaction time for data taken using linearly decreasing (panels b and f), random (panels c and g) and increasing (panels d and h) sequences of magnetic field. Pulse width

= 10 μs. Gradient length = 1050 μs. Diffusion time = 50 ms. Relaxation delay time = 3 s (panels b and f). Pulse width = 10 μs. Gradient length = 1100 μs. Diffusion time = 50 ms. Relaxation delay time = 5 s (panels c and g). Reaction conditions: 250 mM reactant, 20 mM catalyst.

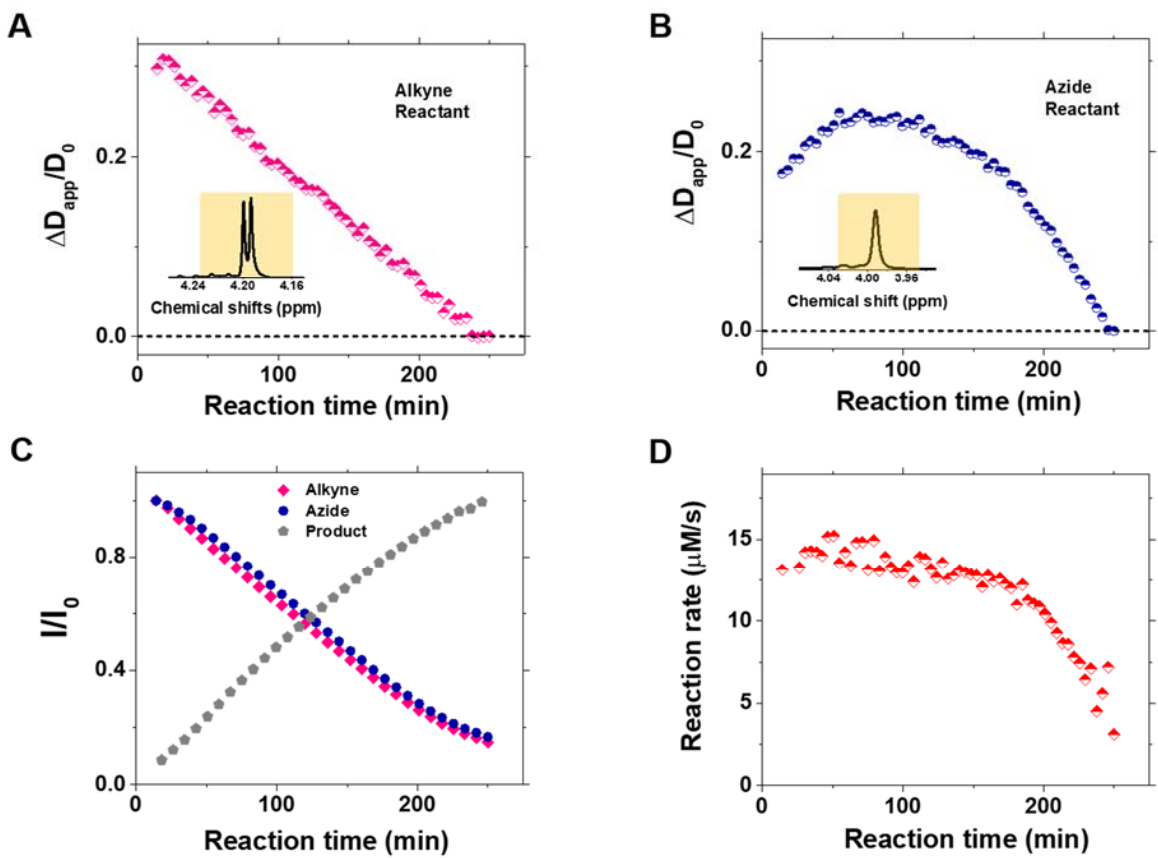

**Figure 3. Diffusion data for the click reaction with slower kinetic conditions closer to those in ref. 29.** Increased diffusion plotted against reaction time for the alkyne reactant (**A**) and the azide reactant (**B**) with respective NMR peaks shown as insets. The time-dependent relative concentrations of reactant and product (**C**) and the time-dependent reaction rate (**D**) appear to be close to those asserted in ref. 29. Pulse width = 9.65 μs. Gradient length = 1100 μs. Diffusion time = 50 ms. Relaxation delay time = 10 s. Reaction conditions: 250 mM reactant, 17.5 mM catalyst. The PGSE NMR spectra were acquired using random pulse sequence as recommended in ref. 29.

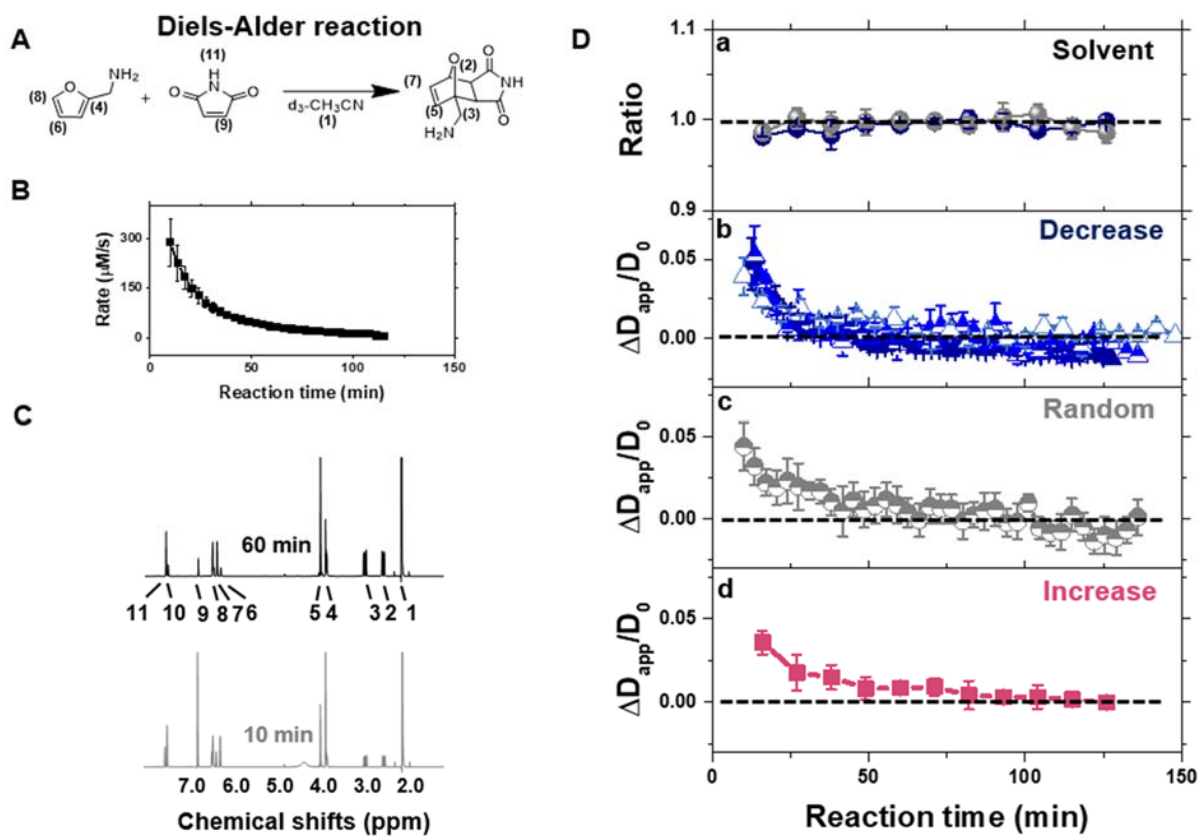

**Figure 4. Diffusion data for the Diels-Alder reaction compared using linearly-increasing, linearly decreasing, and random sequences of magnetic fields.** **(A)** Chemistry of this reaction. **(B)** Kinetics of this reaction, showing smoothed data for the conditions of ref.[27] **(C)** NMR spectra near the start (10 min) and end (60 min) of reaction, showing the NMR peak assignments. **(D)** Comparisons of diffusion measurements of solvent peak (1 in C) from three methods. **(a)** Normalized to data obtained with linear-increasing magnetic field gradient, diffusion coefficients measured with random (circles) and linearly-decreasing (triangles) magnetic field sequences are plotted against reaction time. **(b)** Normalized to $D_o$ at the end of reaction, diffusion coefficient measured with linearly-decreasing magnetic field gradient is plotted against reaction time. Delay between scans is varied: 3s (open symbols), 5s (half-filled symbols), 10s (filled symbols). **(c)**

Normalized to $D_o$ at the end of reaction, diffusion coefficient measured with random sequence of magnetic field gradient is plotted against reaction time. **(d)** Normalized to $D_o$ at the end of reaction, diffusion coefficient measured with linearly-increasing magnetic field gradient is plotted against reaction time (this data is taken from ref.[27]).

# REFERENCES

1. Ramaswamy, S., Active matter. *J. Stat. Mech.: Theory Exp* **2017**, *2017* (5), 054002.

2. Marchetti, M. C.; Joanny, J. F.; Ramaswamy, S.; Liverpool, T. B.; Prost, J.; Rao, M.; Simha, R. A., Hydrodynamics of soft active matter. *Rev. Mod. Phys.* **2013**, *85* (3), 1143-1189.

3. Jee, A.-Y.; Cho, Y.-K.; Granick, S.; Tlusty, T., Catalytic enzymes are active matter. *Proc. Natl. Acad. Sci. U.S.A* **2018**, *115* (46), E10812-E10821.

4. Kassem, S.; van Leeuwen, T.; Lubbe, A. S.; Wilson, M. R.; Feringa, B. L.; Leigh, D. A., Artificial molecular motors. *Chem. Soc. Rev.* **2017**, *46* (9), 2592-2621.

5. Liebchen, B.; Löwen, H., Synthetic chemotaxis and collective behavior in active matter. *Acc. Chem. Res.* **2018**, *51* (12), 2982-2990.

6. Bechinger, C.; Di Leonardo, R.; Löwen, H.; Reichhardt, C.; Volpe, G.; Volpe, G., Active particles in complex and crowded environments. *Rev. Mod. Phys.* **2016**, *88* (4), 045006.

7. Garcia-Lopez, V.; Chen, F.; Nilewski, L. G.; Duret, G.; Aliyan, A.; Kolomeisky, A. B.; Robinson, J. T.; Wang, G.; Pal, R.; Tour, J. M., Molecular machines open cell membranes. *Nature* **2017**, *548* (7669), 567-572.

8. Ghosh, S.; Somasundar, A.; Sen, A., Enzymes as active matter. *Annu. Rev. Condens. Matter Phys.* **2021**, *12* (1), 177-200.

9. Pressé, S., A thermodynamic perspective on enhanced enzyme diffusion. *Proc. Natl. Acad. Sci. U. S. A.* **2020**, 202022207.

10. Antalek, B., Using pulsed gradient spin echo NMR for chemical mixture analysis: How to obtain optimum results. *Concept Magn. Reson. B* **2002**, *14* (4), 225-258.

11. Jr., C. S. J., Diffusion ordered nuclear magnetic resonance spectroscopy: principles and applications. *Prog. Nucl. Mag. Res. Spectrosc.* **1999**, *34*.

12. Tanner, J. E., Use of the stimulated echo in NMR diffusion studies. *J. Chem. Phys.* **1970**, *52* (5), 2523-2526.

13. Stejskal, E. O.; Tanner, J. E., Spin diffusion measurements: spin echoes in the presence of a time‐dependent field gradient. *J. Chem. Phys.* **1965**, *42* (1), 288-292.

14. Woessner, D. E., Effects of diffusion in nuclear magnetic resonance spin‐echo experiments. *J. Chem. Phys.* **1961**, *34* (6), 2057-2061.

15. Oikonomou, M.; Asencio-Hernandez, J.; Velders, A. H.; Delsuc, M. A., Accurate DOSY measure for out-of-equilibrium systems using permutated DOSY (p-DOSY). *J. Magn. Reson.* **2015**, *258*, 12-6.

16. Khajeh, M.; Botana, A.; Bernstein, M. A.; Nilsson, M.; Morris, G. A., Reaction kinetics studied using Diffusion-Ordered Spectroscopy and multiway chemometrics. *Anal. Chem.* **2010**, *82* (5), 2102-2108.

17. Nilsson, M.; Khajeh, M.; Botana, A.; Bernstein, M. A.; Morris, G. A., Diffusion NMR and trilinear analysis in the study of reaction kinetics. *Chem. Commun.* **2009**, (10), 1252-4.

18. Li, D.; Keresztes, I.; Hopson, R.; Williard, P. G., Characterization of reactive intermediates by multinuclear Diffusion-Ordered NMR Spectroscopy (DOSY). *Acc. Chem. Res.* **2009**, *42* (2), 270-280.

19. Cabrita, E. J.; Berger, S.; Brauer, P.; Karger, J., High-resolution DOSY NMR with spins in different chemical surroundings: influence of particle exchange. *J Magn. Reson.* **2002**, *157* (1), 124-31.

20. Jee, A.-Y.; Chen, K.; Tlusty, T.; Zhao, J.; Granick, S., Enhanced diffusion and oligomeric enzyme dissociation. *J. Am. Chem. Soc.* **2019**, *141* (51), 20062-20068.

21. Zhang, Y.; Armstrong, M. J.; Bassir Kazeruni, N. M.; Hess, H., Aldolase does not show enhanced diffusion in dynamic light scattering experiments. *Nano Lett.* **2018**, *18* (12), 8025-8029.

22. Gounarides, J. S.; Chen, A.; Shapiro, M. J., Nuclear magnetic resonance chromatography: applications of pulse field gradient diffusion NMR to mixture analysis and ligand–receptor interactions. *J. Chromatogr. B Biomed. Appl.* **1999**, *725* (1), 79-90.

23. Viel, S.; Capitani, D.; Mannina, L.; Segre, A., Diffusion-ordered NMR spectroscopy: a versatile tool for the molecular weight determination of uncharged polysaccharides. *Biomacromolecules* **2003**, *4* (6), 1843-1847.

24. Groves, P., Diffusion ordered spectroscopy (DOSY) as applied to polymers. *Poly. Chem.* **2017**, *8* (44), 6700-6708.

25. Zuccaccia, C.; Stahl, N. G.; Macchioni, A.; Chen, M.-C.; Roberts, J. A.; Marks, T. J., NOE and PGSE NMR spectroscopic studies of solution structure and aggregation in metallocenium ion-pairs. *J. Am. Chem. Soc.* **2004**, *126* (5), 1448-1464.

26. Song, F.; Lancaster, S. J.; Cannon, R. D.; Schormann, M.; Humphrey, S. M.; Zuccaccia, C.; Macchioni, A.; Bochmann, M., Synthesis, ion aggregation, alkyl bonding modes, and dynamics of 14-electron metallocenium ion pairs [(SBI)MCH2SiMe3+···X-] (M = Zr, Hf): inner-sphere (X = MeB(C6F5)3) versus outer-sphere (X = B(C6F5)4) structures and the implications for "continuous" or "intermittent" alkene polymerization mechanisms. *Organometallics* **2005,** *24* (6), 1315-1328.

27.     Wang, H.; Park, M.; Dong, R.; Kim, J.; Cho, Y.-K.; Tlusty, T.; Granick, S., Boosted molecular mobility during common chemical reactions. *Science* **2020,** *369* (6503), 537-541.

28.     Wang, H.; Park, M.; Dong, R.; Kim, J.; Cho, Y.-K.; Tlusty, T.; Granick, S., Response to Comment on "Boosted molecular mobility during common chemical reactions". *Science* **2021,** *371* (6526), eabe8678.

29.     Günther, J.-P.; Fillbrook, L. L.; MacDonald, T. S. C.; Majer, G.; Price, W. S.; Fischer, P.; Beves, J. E., Comment on "Boosted molecular mobility during common chemical reactions". *Science* **2021,** *371* (6526), eabe8322.

30.     Swan, I.; Reid, M.; Howe, P. W. A.; Connell, M. A.; Nilsson, M.; Moore, M. A.; Morris, G. A., Sample convection in liquid-state NMR: Why it is always with us, and what we can do about it. *J. Magn. Reson.* **2015,** *252*, 120-129.

31.     Sørland, G. H.; Seland, J. G.; Krane, J.; Anthonsen, H. W., Improved convection compensating pulsed field gradient spin-echo and stimulated-echo methods. *J. Magn. Reson.* **2000,** *142* (2), 323-325.

32.     Data repository-DOI: 10.5281/zenodo.4515126.

33.     Hoffman, R. E.; Arzuan, H.; Pemberton, C.; Aserin, A.; Garti, N., High-resolution NMR "chromatography" using a liquids spectrometer. *J. Magn．Reson．* **2008,** *194* (2), 295-299.

34.     Kiraly, P.; Swan, I.; Nilsson, M.; Morris, G. A., Improving accuracy in DOSY and diffusion measurements using triaxial field gradients. *J. Magn. Reson.* **2016,** *270*, 24-30.

35.     Urbanczyk, M.; Bernin, D.; Czuron, A.; Kazimierczuk, K., Monitoring polydispersity by NMR diffusometry with tailored norm regularisation and moving-frame processing. *Analyst* **2016,** *141* (5), 1745-52.

36.     Cherni, A.; Chouzenoux, E.; Delsuc, M. A., PALMA, an improved algorithm for DOSY signal processing. *Analyst* **2017,** *142* (5), 772-779.

37.     Hamdoun, G.; Guduff, L.; van Heijenoort, C.; Bour, C.; Gandon, V.; Dumez, J. N., Spatially encoded diffusion-ordered NMR spectroscopy of reaction mixtures in organic solvents. *Analyst* **2018,** *143* (14), 3458-3464.

38.     Price, W. S.; Kuchel, P. W., Effect of nonrectangular field gradient pulses in the stejskal and tanner (diffusion) pulse sequence. *J. Magn. Reson.* **1991,** *94* (1), 133-139.

39.     Jehenson, P.; Westphal, M.; Schuff, N., Analytical method for the compensation of eddy-current effects induced by pulsed magnetic field gradients in NMR systems. *J. Magn. Reson.* **1990,** *90* (2), 264-278.

40.     Håkansson, B.; Jönsson, B.; Linse, P.; Söderman, O., The influence of a nonconstant magnetic-field gradient on PFG NMR diffusion experiments. a Brownian-dynamics computer simulation study. *J. Magn. Reson.* **1997,** *124* (2), 343-351.

41.     Freeman, R., Shaped radiofrequency pulses in high resolution NMR. *Prog. Nucl. Mag. Res. Spectrosc.* **1998,** *32* (1), 59-106.

42.     Goux, W. J.; Verkruyse, L. A.; Saltert, S. J., The impact of Rayleigh-Benard convection on NMR pulsed-field-gradient diffusion measurements. *J. Magn. Reson.* **1990,** *88* (3), 609-614.

43.     Mao, X.-A.; Kohlmann, O., Diffusion-broadened velocity spectra of convection in variable-temperature BP-LED experiments. *J. Magn. Reson.* **2001,** *150* (1), 35-38.

44.     MacDonald, T. S. C.; Price, W. S.; Astumian, R. D.; Beves, J. E., Enhanced diffusion of molecular catalysts is due to convection. *Angew. Chem. Int. Ed.* **2019,** *131* (52), 19040-19043.

45.     Xie, J.; Hase, W. L., Rethinking the $S_N2$ reaction. *Science* **2016,** *352* (6281), 32-33.

**TOC Graphic**

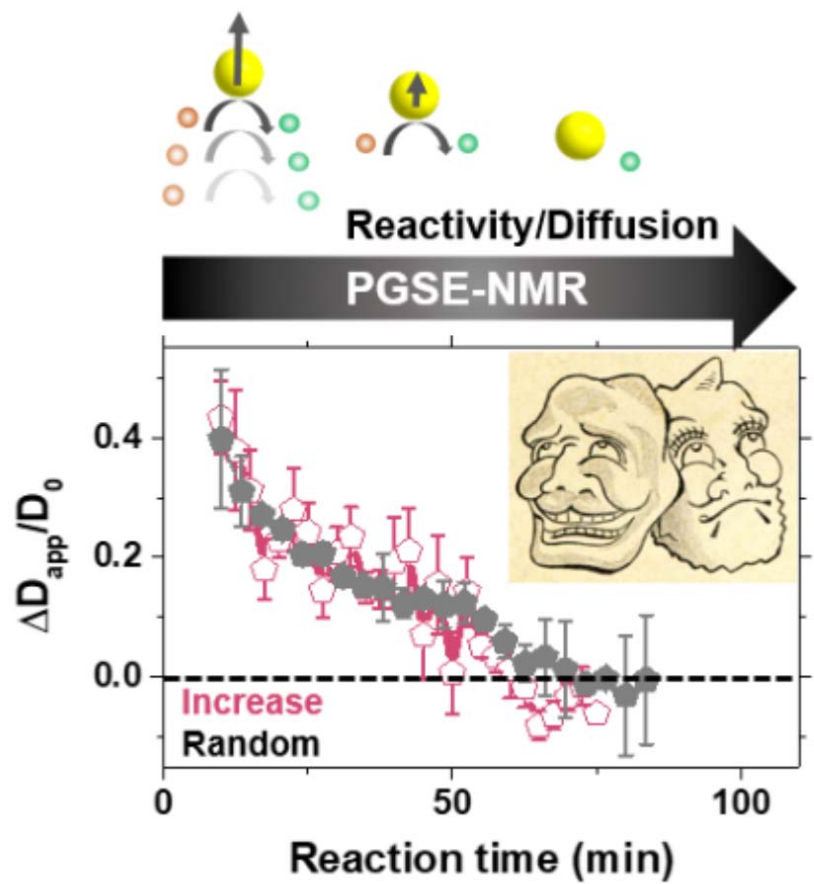